\documentclass[prb,twocolumn,showpacs,floatfix]{revtex4-1}

\usepackage[utf8]{inputenc}
\usepackage{graphicx}
\usepackage{flafter}
\usepackage{amsmath}
\usepackage{amssymb}
\usepackage{float}
\usepackage{color}
\usepackage{subfigure}

\begin{document}
\title{Scattering of two-dimensional Dirac fermions on gate-defined oscillating quantum dots}
\author{C. Schulz}
\affiliation{Institut f{\"u}r Physik,
Ernst-Moritz-Arndt-Universit{\"a}t Greifswald, 17487 Greifswald, Germany }
\author{R. L. Heinisch}
%\altaffiliation[]
%{Present address: Ausw\"artiges Amt, Werderscher Markt 1, 10117 Berlin, Germany}
\affiliation{Institut f{\"u}r Physik,
Ernst-Moritz-Arndt-Universit{\"a}t Greifswald, 17487 Greifswald, Germany }
\author{H. Fehske}
\affiliation{Institut f{\"u}r Physik,
Ernst-Moritz-Arndt-Universit{\"a}t Greifswald, 17487 Greifswald, Germany }
\date{\today}
\begin{abstract}
Within an effective Dirac-Weyl  theory we solve the scattering problem for massless chiral fermions impinging  on a cylindrical  time-dependent potential barrier.  
The set-up we consider can be used to model the electron propagation in a monolayer of graphene with harmonically driven quantum dots.  For static small-sized quantum dots scattering resonances enable particle confinement and interference effects may switch forward scattering  on and off. An oscillating  dot may cause inelastic scattering by excitation of  states with energies shifted by integer multiples of the oscillation frequency, which  significantly modifies the scattering characteristics of 
static dots. Exemplarily the scattering  efficiency of a  potential barrier with zero bias remains finite in the limit of low particle energies and small  potential amplitudes.  For an oscillating  quantum dot with finite bias, the partial wave resonances at higher energies are smeared out for small frequencies or large oscillation amplitudes, thereby dissolving  the quasi-bound states at the quantum dot.    
\end{abstract}


\maketitle

\section{Introduction}
Graphene-based nanostructures feature striking and sometimes counter-intuitive transport properties that  primarily arise from the linear form of the (gapless) energy spectrum near the so-called Dirac nodal points and the related nontrivial topology of the wave function~\cite{CGPNG09}. One consequence of the pseudo-relativistic dynamics of such massless chiral Dirac-Weyl quasiparticles~\cite{Di28,We29}, having an additional  pseudospin degree of freedom, is their perfect transmission through arbitrarily high and wide rectangular potential barriers or $n$-$p$ junctions at perpendicular incidence. Recent experiments~\cite{SHG09,NKPKKL11} confirm this so-called Klein tunneling phenomenon~\cite{Kl28,KNG06}  which seems to prevent an electrostatic confinement of Dirac electrons.
   
For large circular $n$-$p$ junctions, on the other hand,  refraction gives rise to two caustics that coalesce in a cusp and therefore focusses the particle density inside the gated region~\cite{CPP07,AU13,WF14}.  
Resonances in the conductance~\cite{BTB09,PAS11} and the scattering cross section~\cite{HBF13a}  indicate the formation of quasi-bound electron states 
also for small circular gate-defined quantum dots in monolayer graphene. Thereby, forward scattering and Klein tunneling can be almost switched off by a Fano resonance
arising from the interference between resonant scattering and the background partition~\cite{HBF13a}.  
For the density of states the presence of well-quantized states in the quantum dot leads to an additional peak structure~\cite{SB14}. These results, obtained within Dirac theory, were confirmed for a 
tight-binding  graphene lattice model utilizing exact numerical techniques~\cite{PHF13}. From an application-oriented point of view, graphene quantum dots with `confined' 
electrons may serve as hosts for spin qbits~\cite{PSKYHNG08,WRAWRB09,RT10}. 

Applying time-dependent external fields to graphene nanostructures or mesoscopic devices may lead to new transport phenomena. For example, the relevance of photon-assisted transport~\cite{PA04} to the observability of {\it Zitterbewegung} has been shown~\cite{TBM07}. Furthermore, in graphene systems with harmonically time-driven potentials energy-dependent transmission 
and, in particular, inelastic tunneling  can appear where the electrons exchange energy quanta with the oscillating field~\cite{ZST08}. Thereby ,the charge carrier is transferred to electronic side-bands, separated from the particle energy  by multiples of the field modulation frequency,  when passing through the field range. On the other hand,  
electronic transport through a graphene  $n$-$p$ junction irradiated by an electromagnetic field might be suppressed by the creation of a dynamic gap between the electron and hole bands in the quasipartcle spectrum~\cite{FE07}. More recently, for a Dirac electron tunneling problem with a time-dependent potential wall current resonances have been found along the tunnel barrier, which can be resonantly amplified and exhibit a nonzero dc component at specific frequencies, similar to Shapiro steps of driven Josephson junctions~\cite{SHH12}.

Motivated by these multifaceted findings, in the present paper we study the relativistic scattering of Dirac-Weyl particles on a cylindrical time-modulated potential barrier, 
realizing in a plane graphene sheet a single, gate-defined, harmonically driven quantum dot.  In particular, we ask how an oscillating quantum dot affects the electron propagation and backscattering 
 when energy is not conserved.

\section{Theoretical approach}
\subsection{Model}
At low  energies, when the continuum limit and the effective mass approximation applies, the physics of graphene is described by two copies of massless Dirac-like  Hamiltonians, which hold for momenta around the Dirac points $K$ and $K^\prime$ at the corners of the graphene's (hexagonal) first Brillouin zone where the completely filled $\pi$-electron valence   and empty $\pi^\ast$-electron conduction bands  touch~\cite{Wa47}. Near these points both bands have a linear dispersion. In this regime,  the  wave function  obeys the time-independent 2D Dirac-Weyl equation  $-i \hbar v_F \boldsymbol{\sigma \cdot \nabla}\psi({\bf r})=E\psi({\bf r})$, where $E$ is the particle's energy. The vector of Pauli matrices $\boldsymbol{\sigma}=\left(\sigma_{x},\sigma_{y}\right)$,  representing a sublattice pseudospin, acts on the two-component spinor  $\psi({\bf r})$.  Note that the Dirac fermions have opposite helicities in the valleys around $K$ and $K^\prime$,  $\boldsymbol{\sigma \cdot p}/p = +1$ or $-1$; to the valleys $K$ and $K^\prime$ pseudospins $\xi=1$ and -1 can be assigned, respectively.  Obviously the helicity operator commutes with the massless  2D Dirac Hamiltonian, i.e., the helicity---coinciding with the chirality in this  case---is a good quantum number.  Then the calculations  can be carried out for each valley $\xi=1, -1$ separately.   This does not apply, of course,  at larger energies, where the band structure of graphene deviates from the isotropic cone spectrum, or if noticeable intervalley scattering  occurs, e.g., due to strong short-ranged impurity potentials. 

\begin{figure}[t]
\includegraphics[width=8.6cm]{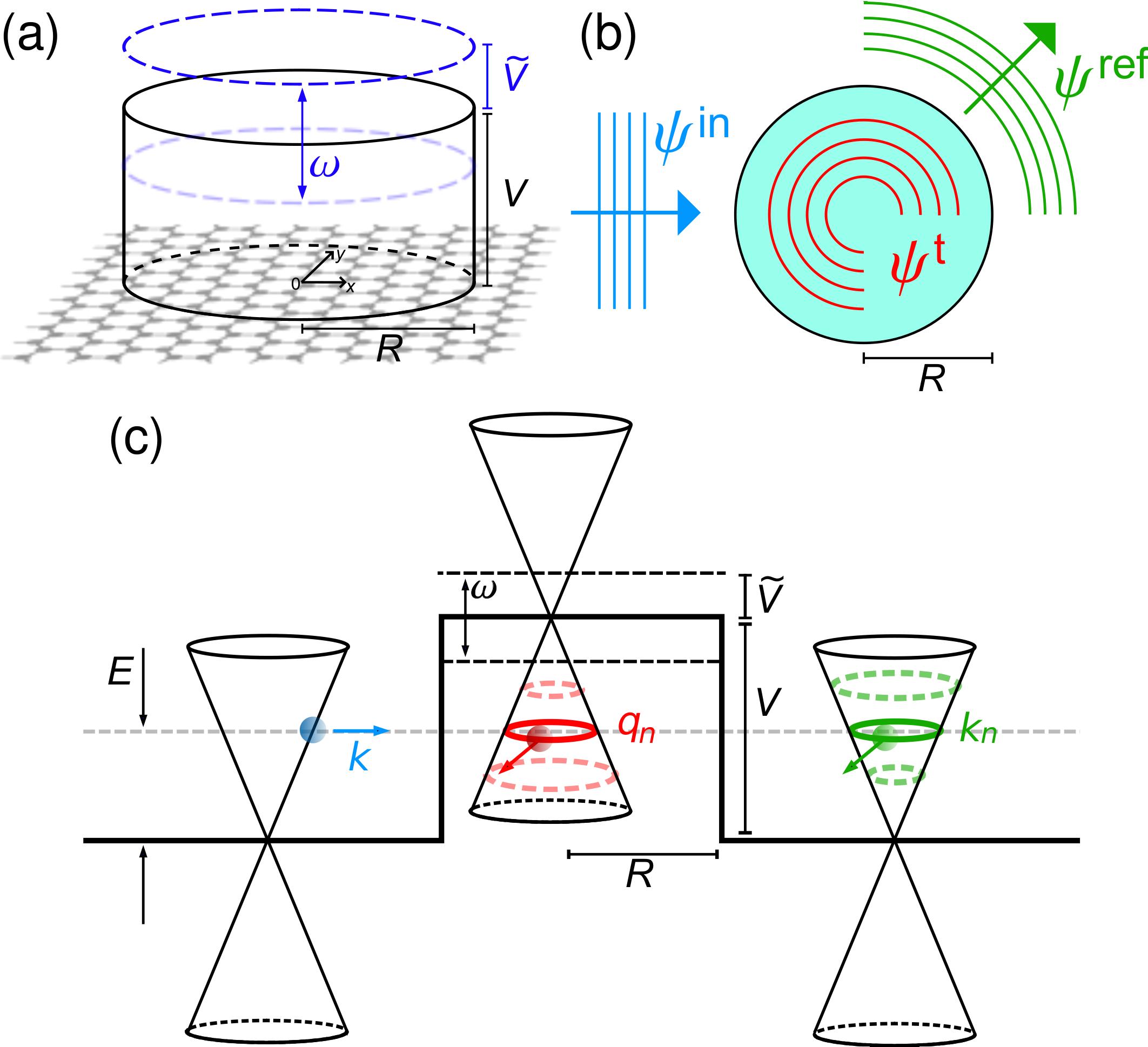}
\caption{(Color online) Sketch of the setup we consider in the present work. A  low-energy plane Dirac electron wave $\psi^{\rm in}$, propagating in a monolayer graphene sheet on a gated substrate, hits a circular  time-dependent potential step that can be tuned by applying a voltage. The  gate-defined graphene quantum dot  is characterized by the constant ($V$) and oscillating ($\tilde{V}$, $\omega$) parts of the potential, and the radius $R$ [see panel (a)]. In the process of scattering reflected $(\psi^{\rm ref})$ and transmitted $(\psi^{\rm t})$ waves appear [panel (b)]. Panel (c) schematically shows the bandstructure  and energy conditions. On account of the time-dependent potential the particle's energy $E$ is not conserved. The reflected and transmitted particles have quantized energies, $E_n=E+n\hbar \omega$ where $n=0, \pm 1, \pm 2,\ldots$ (only the first excited energies were marked in the plot), and carry an angular momentum (i.e., their wave vectors have components in any planar direction).} 
\label{pic0}
\end{figure}

In what follows we focus on electrons in a single graphene  layer subjected to an external scalar potential $U({\bf r}, t)$ that varies slowly in time. 
Neglecting intervalley scattering the low-energy dynamics results from the (single-valley) 
time-dependent Dirac-Weyl equation  
\begin{equation}\label{1}
i \partial_{t}\psi({\bf r}, t)=- i \boldsymbol{\sigma\nabla}\psi({\bf r}, t)+\hat{U}\left(r,t\right)\psi({\bf r}, t) 
\end{equation}
(we use units such that $\hbar=1$, $v_F=1$). Specifically we consider a circular harmonically driven potential step
\begin{equation}\label{2}
\hat{U}\left(r,t\right)=[V+\tilde{V}\sin\omega t]\,\theta\left(R-r\right)\hat{I},
\end{equation}
which is a diagonal operator in spinor space and, in a way, implements a gate-defined quantum dot of radius $R$. Here $V$ is a static barrier and $\tilde{V}$ denotes the amplitude of the potential part that oscillates with angular frequency $\omega$; see  Fig.~\ref{pic0}~(a). Both $V$ and $\tilde{V}$ should vanish outside the gated  region. The use of such a step-like potential---together with the single-valley continuum approximation---has  been justified to a certain extent by the exact numerical treatment of the full (tight-binding model based) scattering problem~\cite{PHF13}, which shows no significant qualitative changes of the scattering behavior when the boundary of the quantum dot was softened adopting a linear interpolation of the potential within a small range $R\pm0.01R$.

\subsection{Solution of the scattering problem}
In the absence of the oscillating potential term, Eq.~\eqref{1} becomes
\begin{equation}\label{3}
[-i\boldsymbol{\sigma\nabla} +V\theta(R-r)\hat{I}\,] \phi (\boldsymbol{r})=E\phi(\boldsymbol{r})\,.
\end{equation}
This eigenvalue problem has been solved before, e.g., in Refs.~\onlinecite{HBF13a,SHF15a}. Using Floquet theory and the Jacobi-Anger identity,  the solution of the time-dependent problem ($\tilde{V},\,\omega\neq 0$) can be constructed in the form $\psi\left(\boldsymbol{r},t\right)=\phi\left(\boldsymbol{r}\right)\chi\left(t\right)$, 
with 
\begin{equation}\label{4}
\chi\left(t\right)=\exp\left(- i Et\right) \sum_{p=-\infty}^\infty i^pJ_p\Big(  \frac{\tilde{V}\theta (R-r)}{\omega} \Big) e^{ip\omega t}\,,
\end{equation}
where $J_p$ denote the Bessel functions of the first kind. According to the scattering geometry displayed in Fig.~\ref{pic0}~(b),
the wave function of the incident electron is assumed to propagate in $x$-direction and can be expanded in polar coordinates ($r,\phi)$:
\begin{eqnarray}\label{5}
\psi^{\text{in}}&=&\frac{1}{\sqrt{2}}{1\choose 1}
e^{i \left(kx-Et\right)} \nonumber \\
&=& \sum_{n}\sum_{m=-\infty}^{\infty}\sqrt{\pi} i^{m+1}\phi_{m,n}^{\left(1\right)}\left(k_{n}r,\varphi\right)\delta_{n0} e^{-i E_{n}t}\,.
\end{eqnarray}
Here $m$  and $n$ are quantum numbers describing the angular momenta and quasi-energies (corresponding to Bloch-Floquet states in a time-dependent periodic potential), respectively. Accordingly the reflected (scattered) and transmitted
waves read: 
\begin{equation}\label{6}
\psi^{\text{ref}}=\sum_{n}\sum_{m=-\infty}^{\infty}\sqrt{\pi} i ^{m+1}\phi_{m,n}^{\left(3\right)}\left(k_{n}r,\varphi\right)r_{m,n} e^{- i E_{n}t}\,,
\end{equation}
\begin{eqnarray}\label{7}
\psi^{\text{t}}&=&\sum_{n}\sum_{m=-\infty}^{\infty}\sqrt{\pi} i^{m+1}\phi_{m,n}^{\left(1\right),\text{t}}\left(q_{n}r,\varphi\right)t_{m,n} e^{-i E_{n}t} \nonumber \\
&& \times \Bigg[\sum_{p=-\infty}^\infty i^p J_p\Big( \frac{\tilde{V}}{\omega} \Big) e^{i p\omega t} \Bigg]
\end{eqnarray}
with scattering coefficients $r_{m,n}$ and $t_{m,n}$. In Eqs.~\eqref{5}-\eqref{7} the eigenfunctions of the Dirac-Weyl Eq.~\eqref{3} are~\cite{HBF13a,SHF15a}
\begin{equation}\label{8}
\phi_{m,n}^{\left(1,3\right)}=\frac{1}{\sqrt{2\pi}}\left(\begin{array}{c}
\begin{array}{c}
- i Z_{m}^{\left(1,3\right)}\left(k_{n}r\right) e^{i m\varphi}\\
\alpha_{n}Z_{m+1}^{\left(1,3\right)}\left(k_{n}r\right) e^{i \left(m+1\right)\varphi}\end{array}\end{array}\right)\,,
\end{equation}
\begin{equation}\label{8b}
\phi_{m,n}^{\left(1\right),\text{t}}=\frac{1}{\sqrt{2\pi}}\left(\begin{array}{c}
\begin{array}{c}
- i Z_{m}^{\left(1\right)}\left(k_{n}r\right) e^{i m\varphi}\\
\alpha_{n}'Z_{m+1}^{\left(1\right)}\left(k_{n}r\right) e^{i\left(m+1\right)\varphi}\end{array}\end{array}\right)\,,
\end{equation}
where $Z_{m}^{\left(1\right)}=J_{m}$, and $Z_{m}^{\left(3\right)}=H_{m}$ are the Hankel's function of the first or second kind. Which kind of Hankel's functions has to be used is determined by the sign of energy: $H_{m}\left(k_{n}r\right)=J_{m}\left(k_{n}r\right)+ i \alpha_{n}Y_{m}\left(k_{n}r\right)$ with `band indices' $\alpha_{n}=\text{sgn}\left(E_{n}\right)$ outside and $\alpha_{n}'=\text{sgn}\left(E_{n}-V\right)$ inside the gated region. In that the scattering is inelastic for our time-dependent Hamiltonian, wave functions with different energies have to be superimposed. The
energy is quantized according to $E_{n}=E+n\omega$ ($n \in \mathbb{Z}$), and the wave numbers $k_{n}=\alpha_{n}E_{n}$ and $q_{n}=\alpha_{n}'\left(E_{n}-V\right)$.
Matching the wave functions at $r=R$, we obtain the following equations of condition for the scattering coefficients:
\begin{equation}\label{9}
\sum_{p=-\infty}^{\infty}t_{m,p} i^{n-p}J_{n-p}\Big(\frac{\tilde{V}}{\omega}\Big)f_{m}^{\left(n,p\right)}=\delta_{n0}g_{m}^{\left(n\right)}\,,
\end{equation}
\begin{eqnarray}\label{10}
r_{m,n}&=&\sum_{p=-\infty}^{\infty}t_{m,p}\frac{J_{m}\left(q_{p}R\right)}{H_{m}\left(k_{n}R\right)} i^{n-p}J_{n-p}\left(\frac{\tilde{V}}{\omega}\right)\nonumber \\
&&-\delta_{n0}\frac{J_{m}\left(k_{n}R\right)}{H_{m}\left(k_{n}R\right)}\,,
\end{eqnarray}
with
\begin{eqnarray}\label{11}
f_{m}^{\left(n,p\right)}&=&H_{m+1}\left(k_{n}R\right)J_{m}\left(q_{p}R\right)\nonumber \\
&&-\alpha_{n}\alpha_{p}'H_{m}\left(k_{n}R\right)J_{m+1}\left(q_{p}R\right)\,,
\end{eqnarray}
\begin{eqnarray}\label{12}
g_{m}^{\left(n\right)}&=&J_{m}\left(k_{n}R\right)H_{m+1}\left(k_{n}R\right)\nonumber \\
&&-\alpha_{n}J_{m+1}\left(k_{n}R\right)H_{m}\left(k_{n}R\right)\,.
\end{eqnarray}
Obviously we have to solve an infinite system of coupled linear equations, whose `coupling strength' is determined by the argument of the Bessel functions $\tilde{V}/\omega$. This has to be done numerically. In doing so, we raise the dimension of the coefficient matrix until convergence is reached, which is most challenging for small values of $\omega$ of course.

The electron density is given by $\rho=\psi^{\dagger}\psi$ and the current by $\boldsymbol{j}=\psi^{\dagger}\boldsymbol{\sigma}\psi$, where $\psi=\psi^{\text{in}}+\psi^{\text{ref}}$ outside and $\psi=\psi^{\text{t}}$ inside the gated dot region. Thereby, the far-field radial component of the reflected current $\psi^{\text{ref}\:\dagger}\boldsymbol{j}\boldsymbol{e_{r}}\psi^{\text{ref}}$,
\begin{eqnarray}\label{13a}
j_{r}^{\text{ref}}\left(r-t,\varphi\right)&=&\sum_{n,p=-\infty}^{\infty}\sum_{m,l=0}^{\infty}\frac{2 e^{i \left(E_{n}-E_{p}\right)\left(r-t\right)}}{\pi r\sqrt{k_{n}k_{p}}} \nonumber \\
&&\hspace*{-2.1cm}\times i^{m-l+\alpha_{p}l-\alpha_{n}m}\left(1-\alpha_{n}i\right)\left(1+\alpha_{p}i\right)r_{m,n}r_{l,p}^{*}\nonumber \\
&&\hspace*{-2.1cm}\times\left[\cos\left(\left(m-l\right)\varphi\right)+\cos\left(\left(m+l+1\right)\varphi\right)\right]\,,
\end{eqnarray}
characterizes the angular scattering. Note that in Eq.~\eqref{13a} and hereafter, $m$ takes non-negative integer values only.   Then the time average of the reflected current,
$\overline{j}_{r}^{\text{ref}}\left(\varphi\right)=\frac{1}{T}\intop_{t}^{t+T}j_{r}^{\text{ref}}\left(t\right)\ \text{d}t$, becomes
\begin{eqnarray}\label{13}
\overline{j}_{r}^{\text{ref}}\left(\varphi\right)&=& \sum_{n=-\infty}^{\infty}\sum_{m,l=0}^{\infty}\frac{4}{\pi rk_{n}}i^{\left(1-\alpha_{n}\right)\left(m-l\right)}r_{m,n}r_{l,n}^{*} \label{13} \\
&& \times\left[\cos\left(\left(m-l\right)\varphi\right)+\cos\left(\left(m+l+1\right)\varphi\right)\right] . \nonumber
\end{eqnarray}

The scattering of a Dirac electron on a circular potential step is advantageously discussed in terms of the scattering efficiency, $Q\left(r-t\right)=\overline{Q}+\tilde{Q}\left(r-t\right)$, that is, the scattering cross section divided by the geometric cross section~\cite{HBF13a}. $Q\left(r-t\right)$ contains two contributions, the time-averaged scattering efficiency, 
\begin{eqnarray}\label{15}
\overline{Q}=\sum_{n=-\infty}^{\infty}\sum_{m=0}^{\infty}\frac{4}{k_{n}R}\left|r_{m,n}\right|^{2} ,
\end{eqnarray}
and a part that is a function of position and time:
\begin{eqnarray}\label{16}
\tilde{Q}\left(r-t\right)&=&\sum_{n<p}\sum_{m=0}^{\infty}\frac{4}{\sqrt{k_{n}k_{p}}R}\Re \Big[e^{ i \left(E_{n}-E_{p}\right)\left(r-t\right)}r_{m,n}r_{m,p}^{*} \nonumber \\
&&\times i^{\left(\alpha_{p}-\alpha_{n}\right)m}\left(1-\alpha_{n} i \right)\left(1+\alpha_{p} i\right)\Big]. \end{eqnarray}

\section{Numerical Results}
The scattering of a plane Dirac electron wave on a cylindrical, electrostatically defined graphene quantum dot with $\tilde{V}=0$ has been analyzed  in previous work~\cite{CPP07,AU13,HBF13a,SHF15a}. In particular, quite recently, the different scattering regimes were classified with regard to the behavior of the cross sections as functions of two parameters that specify the size and the strength of the barrier and also give an estimate of the maximum angular momentum involved in the scattering~\cite{WF14}.  According to this, here we consider basically the quantum domain where the cross sections are determined by resonant scattering. 
In this regime, due to the conical energy dispersion, electrons occupy non-evanescent states inside the dot even when their energy is below the dot potential. For a low energy of the incident electron, scattering resonances due to the excitation of normal modes of the dot appear in distinct preferred scattering directions. At the scattering resonances the electron density in the dot is strongly increased which indicates temporary trapping of the particle. 

In what follows we investigate how the time-dependent modulations of the potential barrier affect the electron propagation. While the scattering at a static quantum dot $(\omega=0)$ is elastic, i.e. the energy is conserved, an oscillating quantum dot can transfer the Dirac particle from the central energy level $E$ to side-bands having  quantized energies $E_n=E+ n\hbar\omega$ with $n=0,\pm 1, \pm 2, \ldots$ 

\subsection{Zero bias}
Let us first consider an oscillating quantum dot with zero bias ($V=0$) near the charge neutrality point $E=0$. In the numerical calculations, we choose---for practical reasons---finite but very small values $V=10^{-9}$ and $E=10^{-10}$. A static quantum dot, of course, will not give any scattering  as $V\to 0$. The harmonically driven quantum dot, on  the other hand, causes reflected and scattered waves due to $\tilde{V},\,\omega>0$. When $V= E\simeq 0$, we have partial waves with $m=0$ only.  
\begin{figure}[t]
\begin{minipage}{4.2cm}
\includegraphics[width=4.8cm,angle=-90]{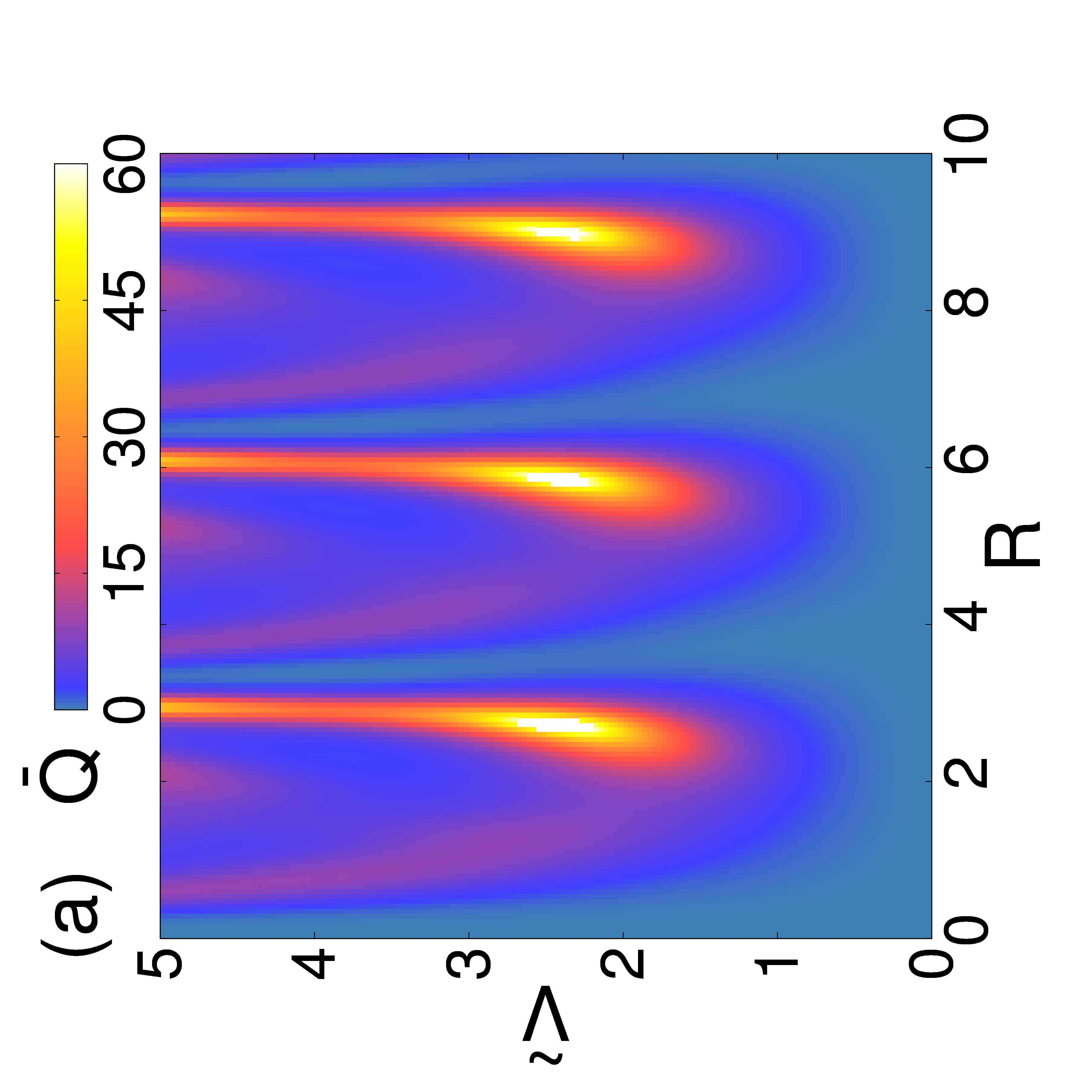}
\end{minipage}
\begin{minipage}{4.2cm}
\includegraphics[width=4.8cm,angle=-90]{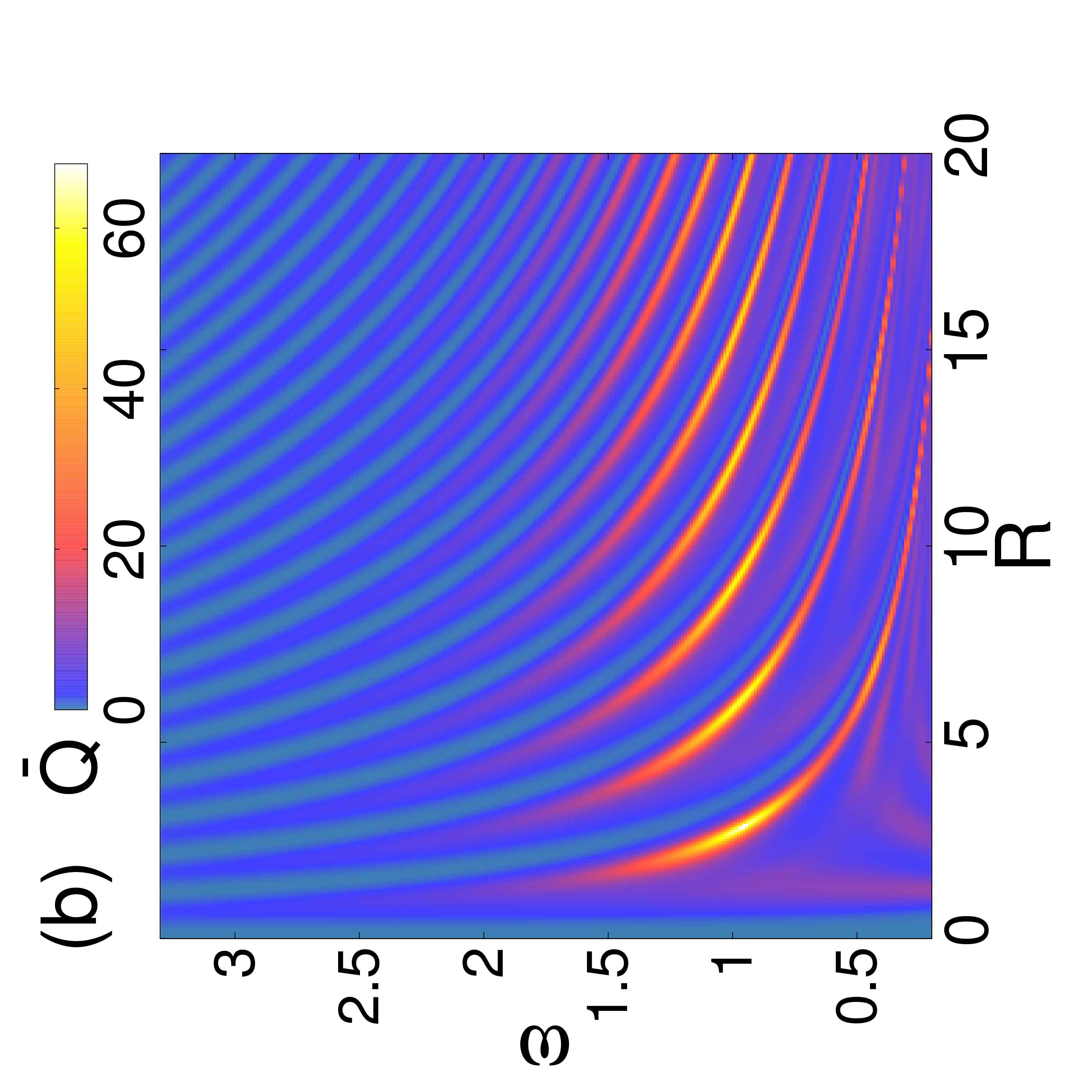}
\end{minipage}\\
\begin{minipage}{4.2cm}
\includegraphics[width=4.cm,angle=0]{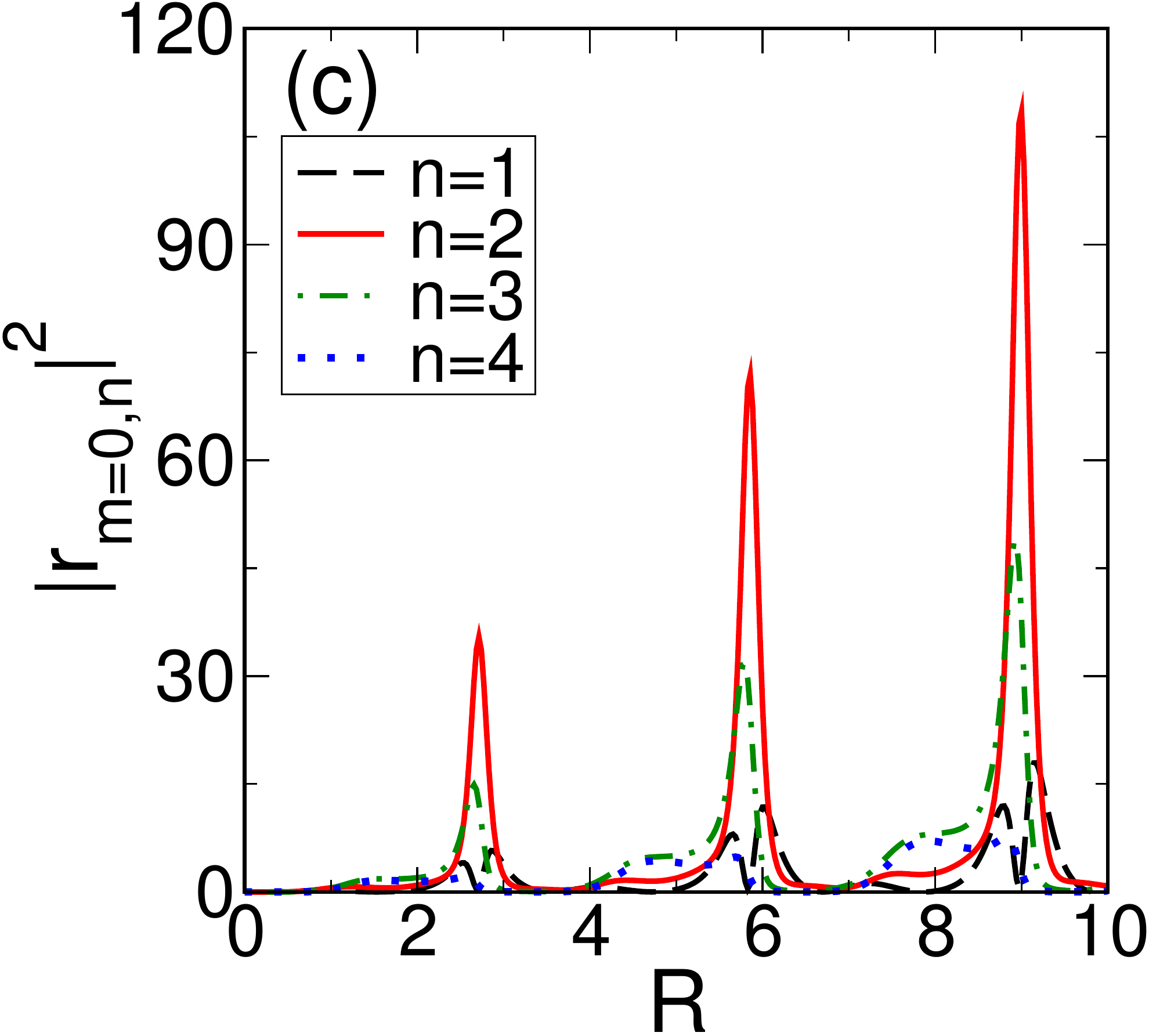}
\end{minipage}
\begin{minipage}{4.2cm}
\includegraphics[width=4.cm,angle=0]{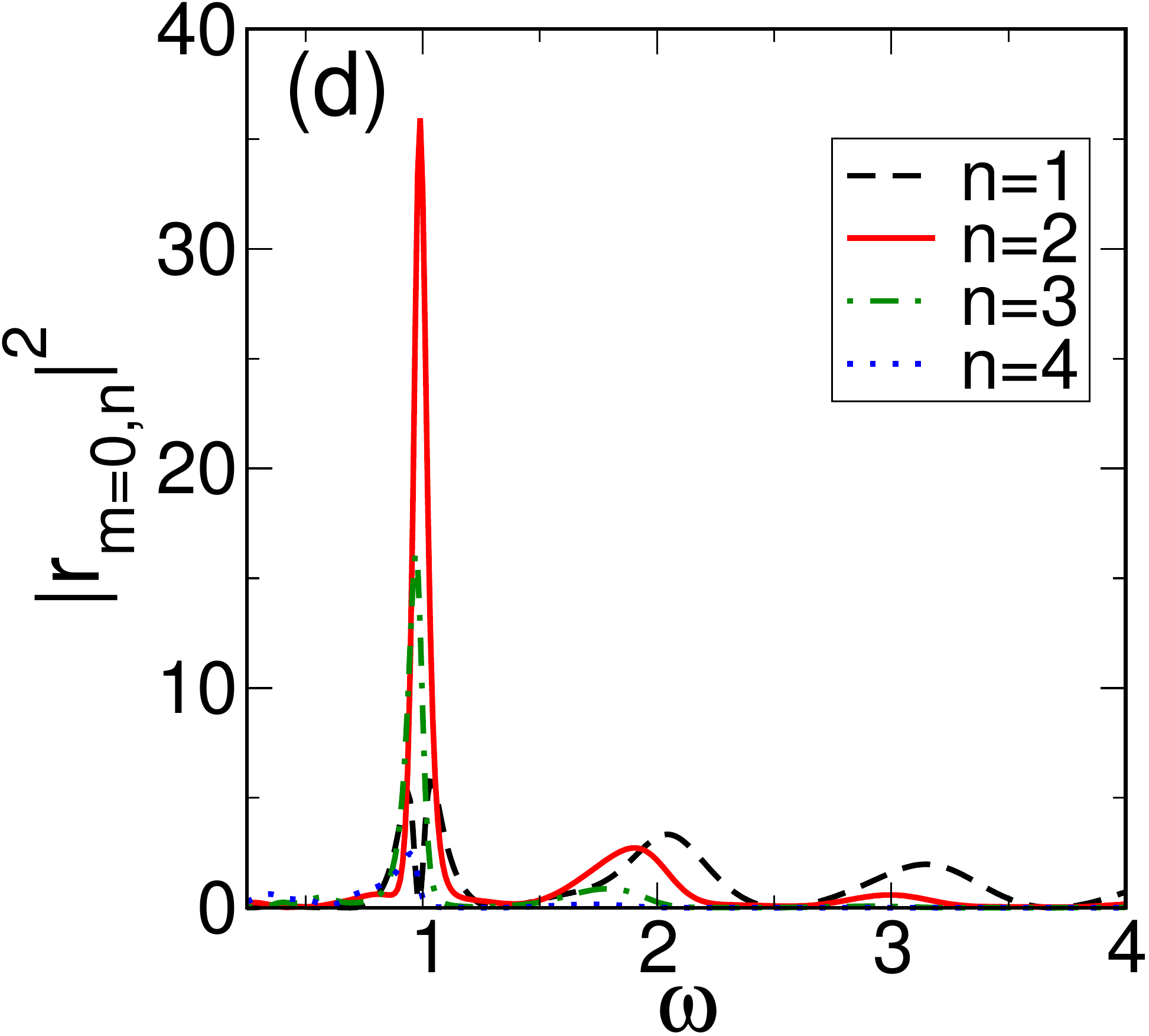}
\end{minipage}
\caption{
\label{pic1}(Color online) Intensity plot of the  time-averaged scattering efficiency $\bar{Q}$ of an oscillating graphene quantum dot with $V=0$ in the far field,  as  function of $R$ and $\tilde{V}$ for $\omega=1$ [panel (a)]  respectively  in dependence on $R$ and $\omega$ for  $\tilde{V}=2.32$ [panel (b)]. 
Panels (c) and (d) give the first four scattering coefficients $|r_{0,n}|^2$ for $\tilde{V}=2.32$, $\omega=1$ and for   $R=2.75$,   $\tilde{V}=2.32$, respectively.}
\end{figure}

Figure~\ref{pic1} presents the far-field scattering efficiency averaged over the time, $\overline{Q}$, in dependence on $R$, $\tilde{V}$ and $\omega$, as well as  
the squared amplitudes of the reflection coefficients Eq.~\eqref{10} contained in  Eq.~\eqref{15}, $|r_{0,n}|^2$, at fixed $\tilde{V}$, $\omega$, respectively, $\tilde{V}$, $R$.
$\overline{Q}$ exhibits a series of pronounced scattering signals at specific parameter ratios of $\tilde{V}$, $\omega$, and $R$; see panels (a) and (b). Basically,   these resonances  can be traced back to quasi-bound states at the quantum dot (which, in a semiclassical picture, correspond to  `standing waves' inside the dot). 
Below the `absorption threshold' of the quantum dot with $V=0$, which is roughly set by $\tilde{V}\simeq \omega=1$ in panel (a), inelastic scattering is negligible.  Panel~(c) indicates that the (non-vanishing) scattering coefficients $|r_{0,n}|^2$ are peaked at the same values of $R$, which are separated by $\Delta R \simeq \pi/\omega$.   
Just above these peaks scattering  is suppressed almost completely. That holds, see panel~(a), even for very large $\tilde{V}$. Hence, varying the frequency 
$\omega$ of the modulation at fixed $R$, $\tilde{V}$, scattering can be turned on and off, which means  that the quantum dot might act as an optical switch [see panel~(b)].  The smaller $\omega$ the more dot eigenmodes can be excited; in particular in the  `adiabatic regime',  $\omega \gtrsim 0$,   scattering  becomes  completely inelastic (cf. also Fig.~\ref{pic4} below).  At large $\omega$, on the other hand,  scattering is exceedingly unlikely and vanishes in the `anti-adiabatic limit' where $\omega\to \infty$, cf. panel (d).

\begin{figure}[t]
\begin{minipage}{4.25cm}
\includegraphics[width=4.9cm,angle=-90]{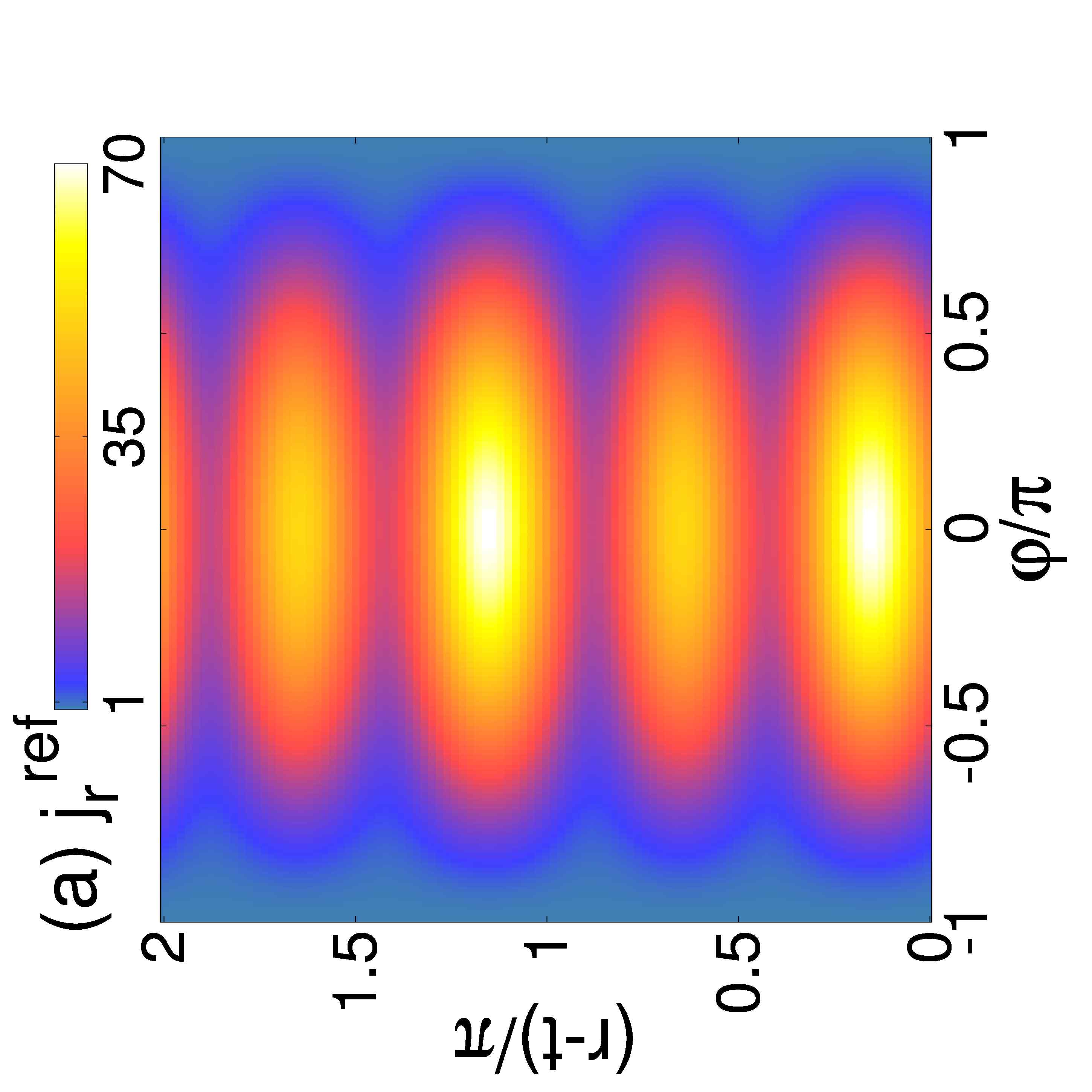}
\end{minipage}
\begin{minipage}{4.25cm}
\vspace{-0.1cm}
\includegraphics[width=4.65cm,angle=-90]{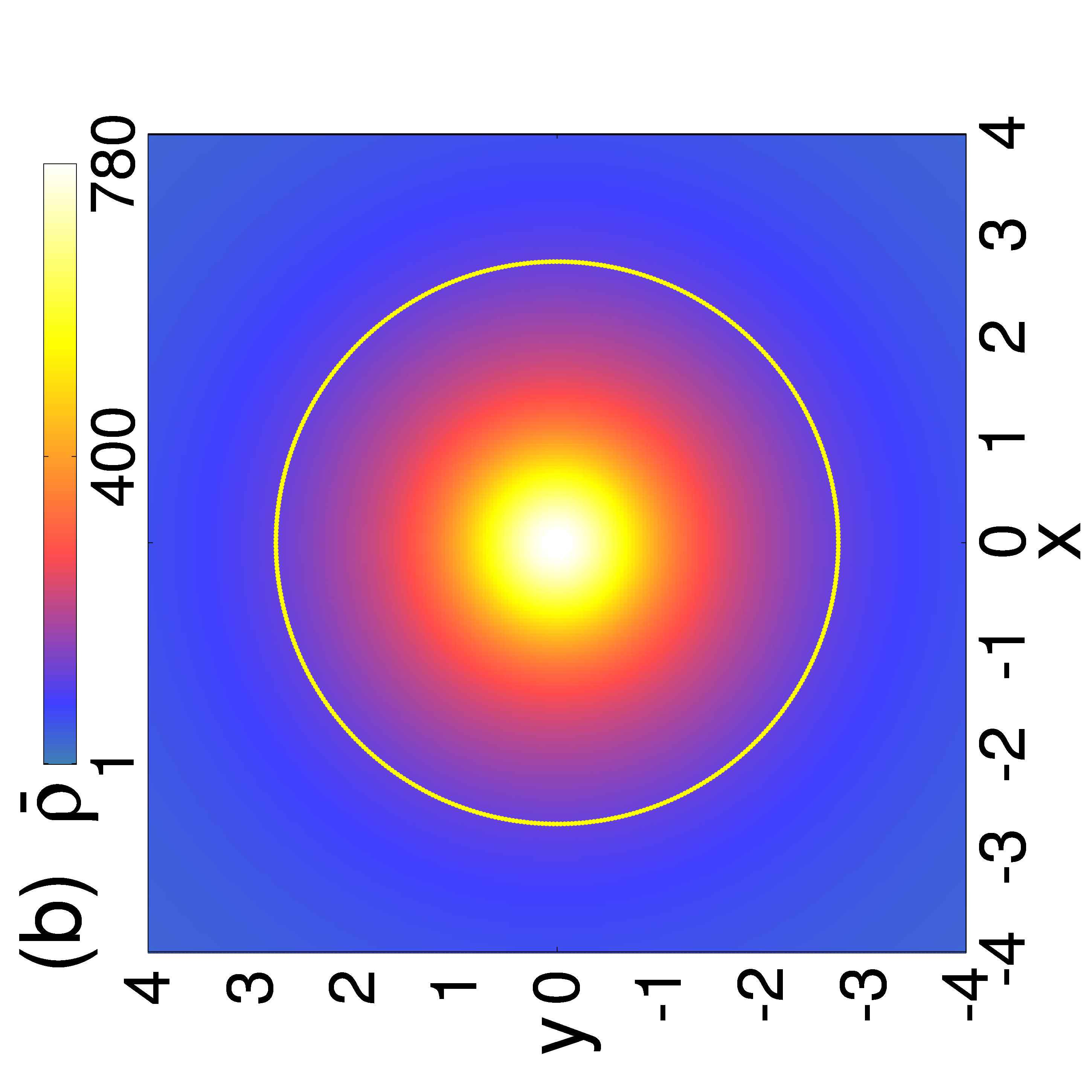}
\end{minipage}
\begin{minipage}{4.25cm}
\includegraphics[width=4.9cm]{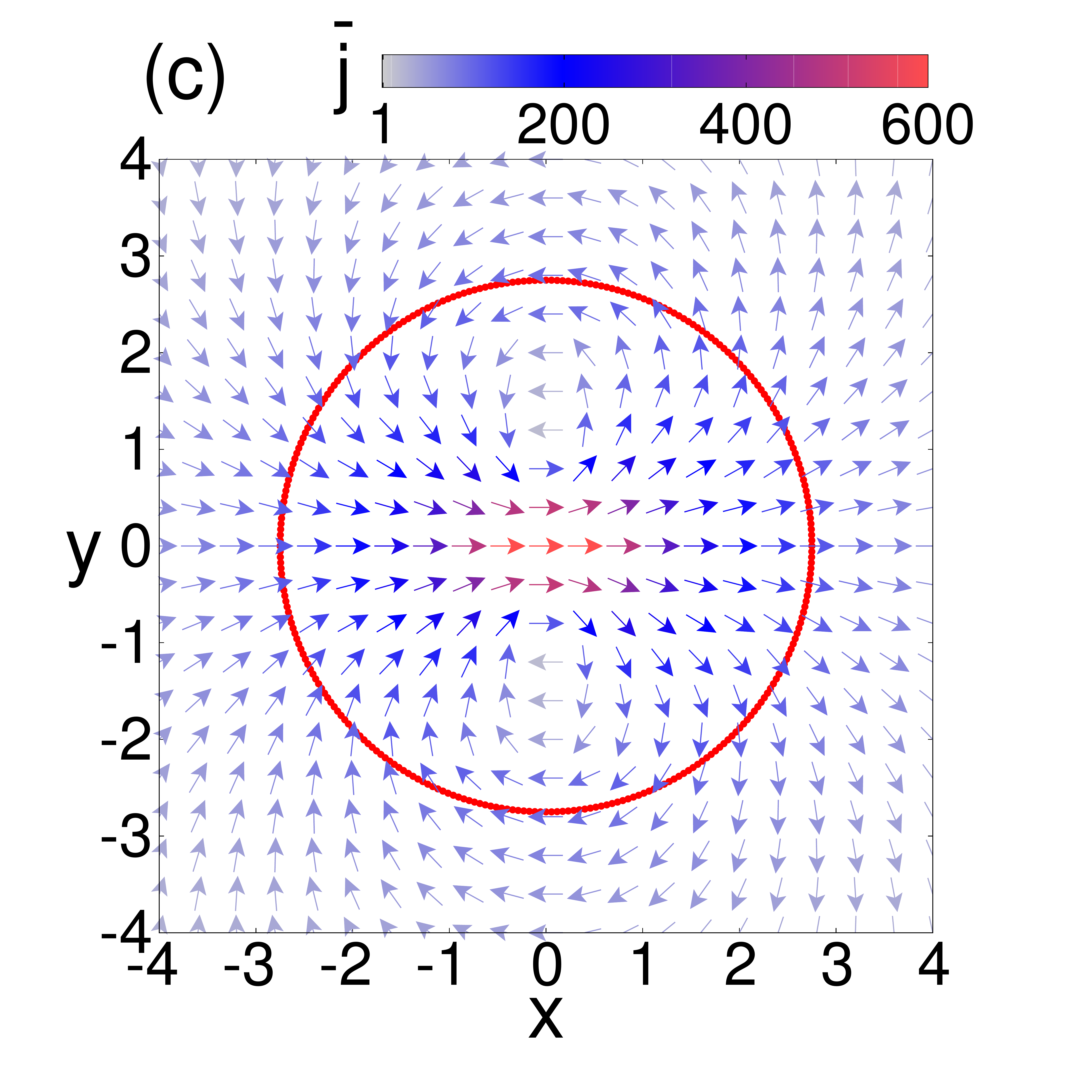}
\end{minipage}
\begin{minipage}{4cm}
\vspace{-0.05cm}
\includegraphics[width=4.03cm]{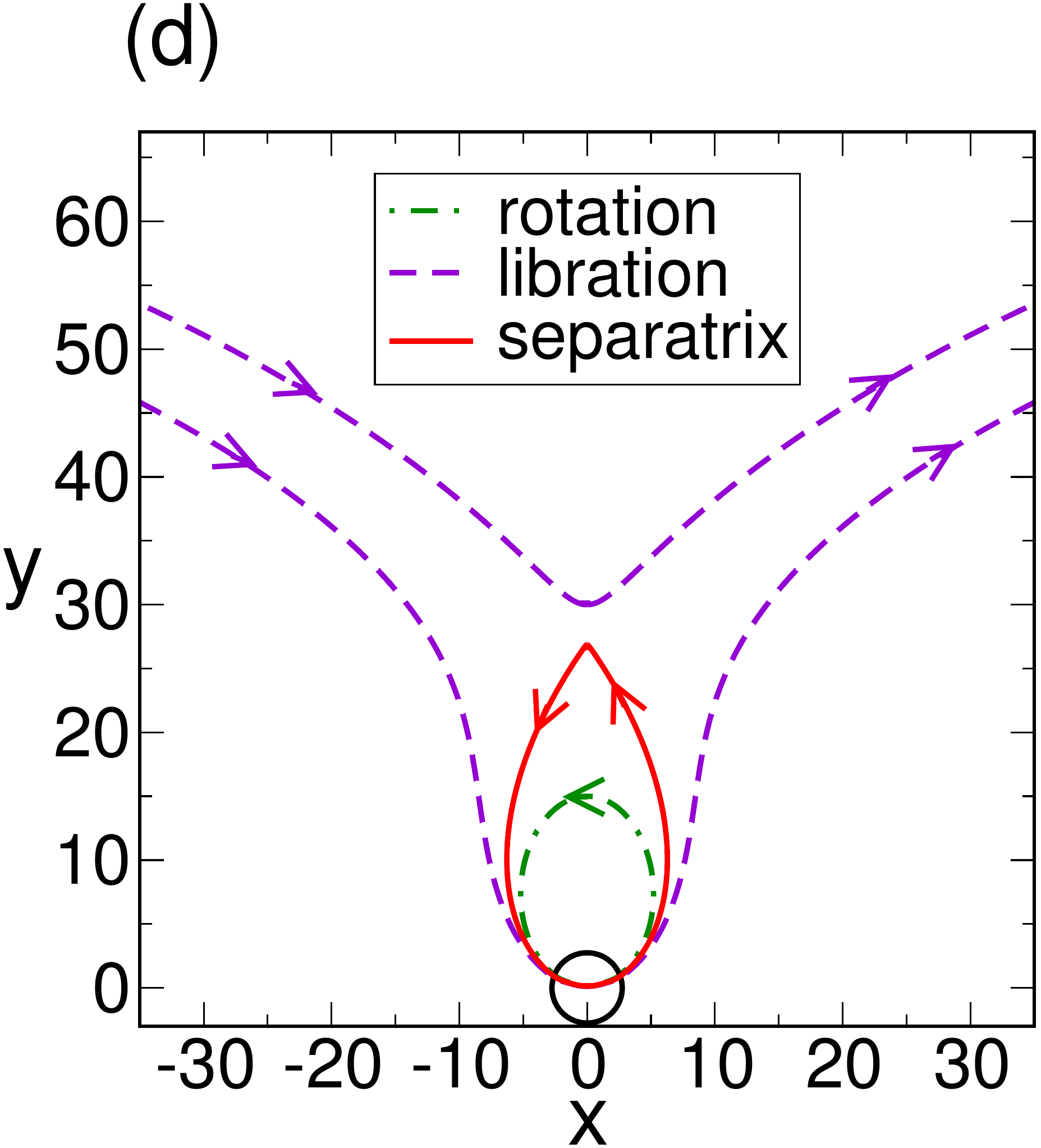}
\end{minipage}
\caption{
\label{pic2}(Color online) Scattering characteristics of an oscillating graphene quantum dot with zero bias. The parameters  $\tilde{V}=2.32$, $\omega=1$, and $R=2.75$ used belong to a strong scattering signal in Fig.~\ref{pic1}~(a). In Fig.~\ref{pic2}, panel (a) displays  the time-dependence of the far-field radial component of the reflected current $j_{r}^{\text{ref}}$ as a function of angle $\varphi$ and phase $(r-t)$.
Panels (b) and (c) show the  time-averaged density $n=\psi^{\dagger}\psi$ and current field $\bar{\boldsymbol{j}}=\psi^{\dagger}\boldsymbol{\sigma}\psi$ 
in the near field, respectively. Panel (d) classifies the current field: unbound librations (violet lines) and bound rotations (green curve) are  separated by a separatrix (red curve). The (yellow, red and black) circles in panels (b), (c) and (d) indicate the spatial extension of the quantum dot.}
\end{figure}

Next we  analyze the angular and temporal dependencies of the radiant emittance by the quantum dot. Because $m=0$ for the oscillating  dot with $V=0$  
forward scattering  should dominate.  This can indeed be seen from the time evolution of the radial component of the reflected current $j_{r}^{\text{ref}}$ in the far-field, depicted in Fig.~\ref{pic2}~(a).  Due to the symmetry  $|r_{0,n}|=|r_{0,-n}| $,  the period of emittance is half the period of the potential oscillation frequency $\omega$. Thereby the incident electron is temporarily captured by the quantum dot and subsequently---as a result of the dot oscillation---reemitted in forward direction. The time-averaged particle density shows that during this process the lowest partial wave becomes resonant, which has maximum electron density in the center of the quantum dot  (at $r=0)$ and leads to a partial trapping of the particle [cf. panel~(b)]. The corresponding pattern of the (near-field) current density is symmetric to the $x$-axis and reveals two vortices where the incident wave is fed into, see panel~(c).  The current field shows that only forward scattering is preferred and Klein tunneling (i.e. perfect transmittance, no backscattering) occurs for those Dirac fermions which perpendicularly impact the boundary of the quantum dot.  The near-field current pattern smoothly turns over into the far-field behavior of the reflected current. Here different regimes  can be distinguished with bound (rotation) and unbound (libration) flow lines separated by a separatrix [cf. panel~(d)]. 

\begin{figure}
\begin{minipage}{4.25cm}
\includegraphics[width=4.9cm,angle=-90]{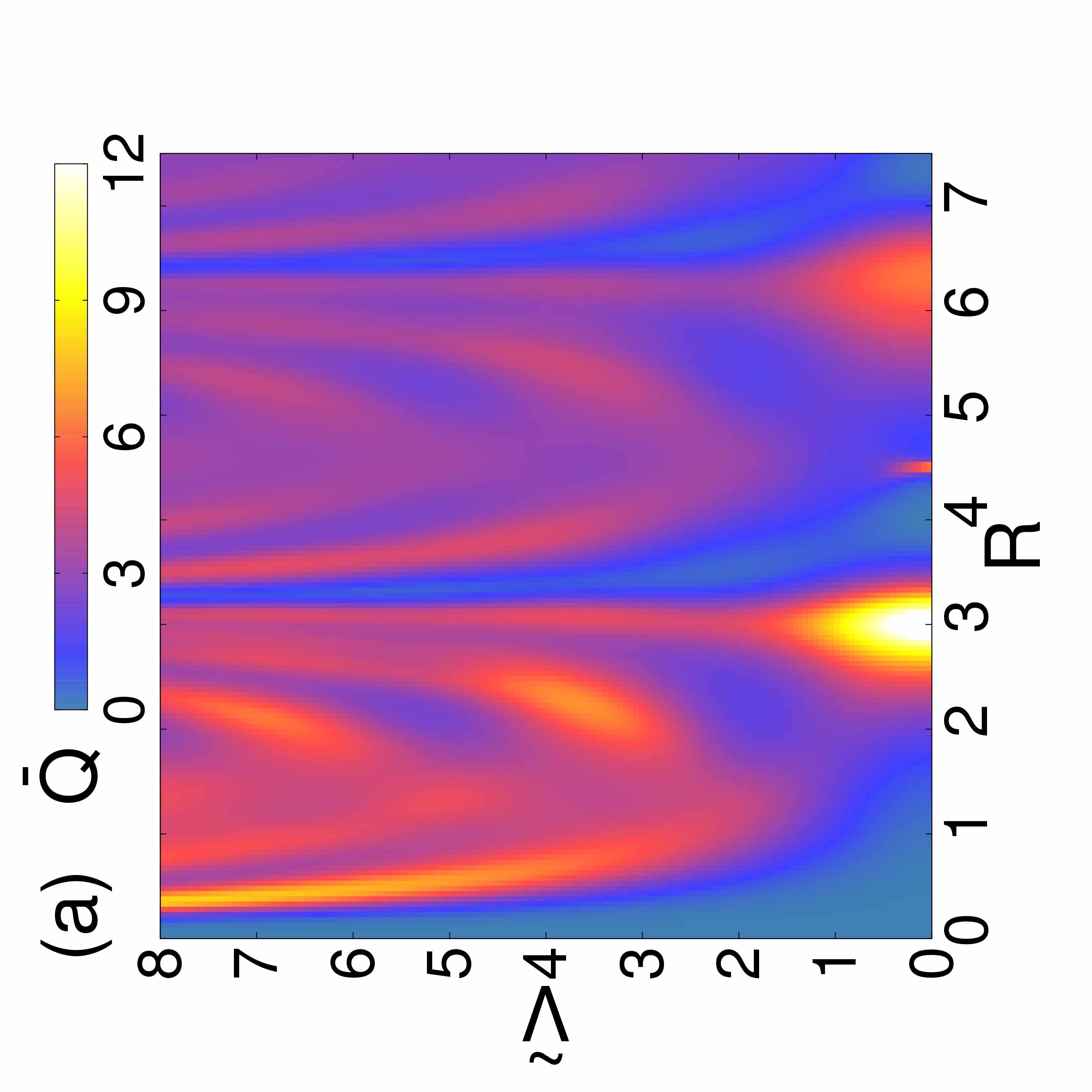}
\end{minipage}
\begin{minipage}{4.25cm}
\includegraphics[width=4.9cm,angle=-90]{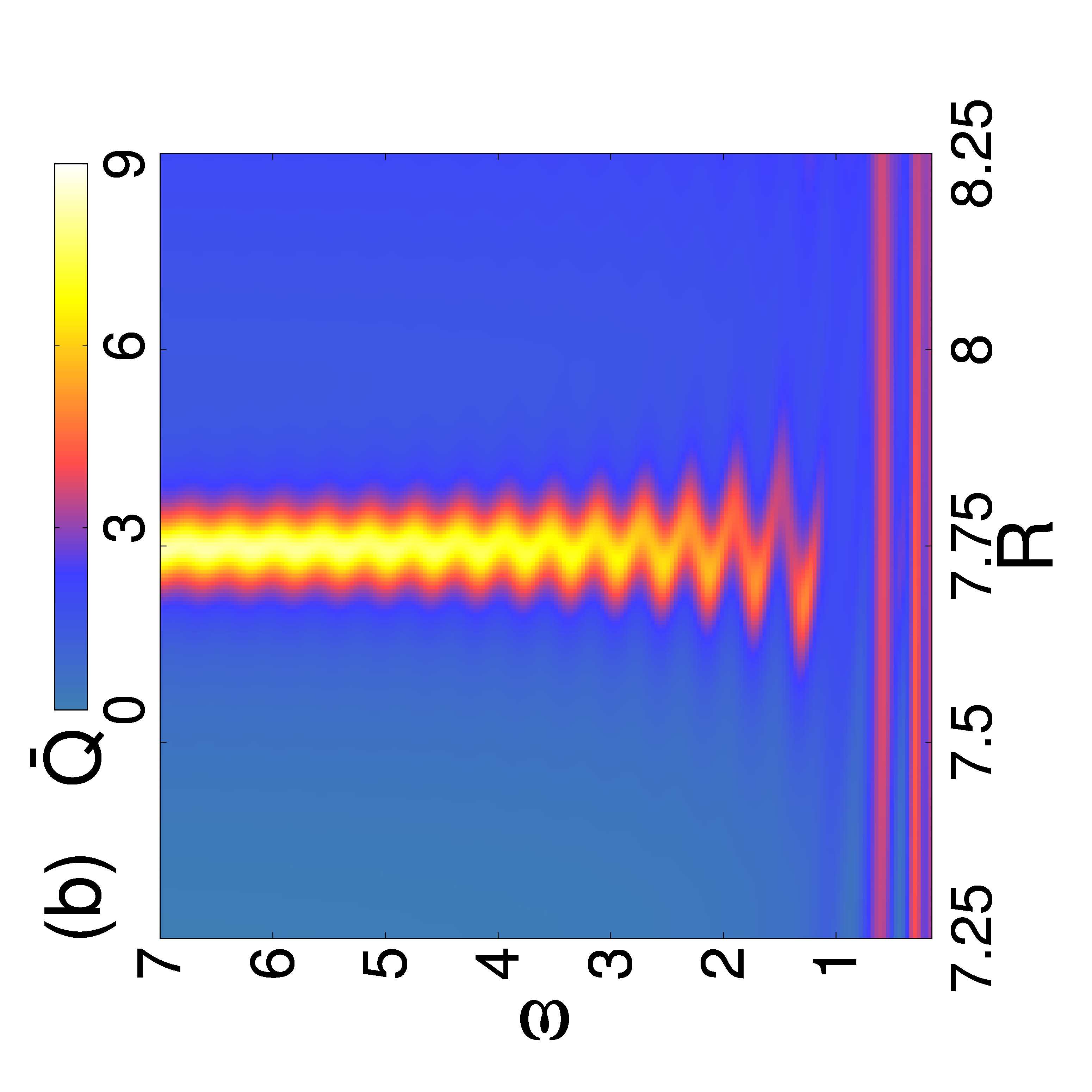}
\end{minipage}
\begin{minipage}{4.25cm}
\includegraphics[width=4.1cm]{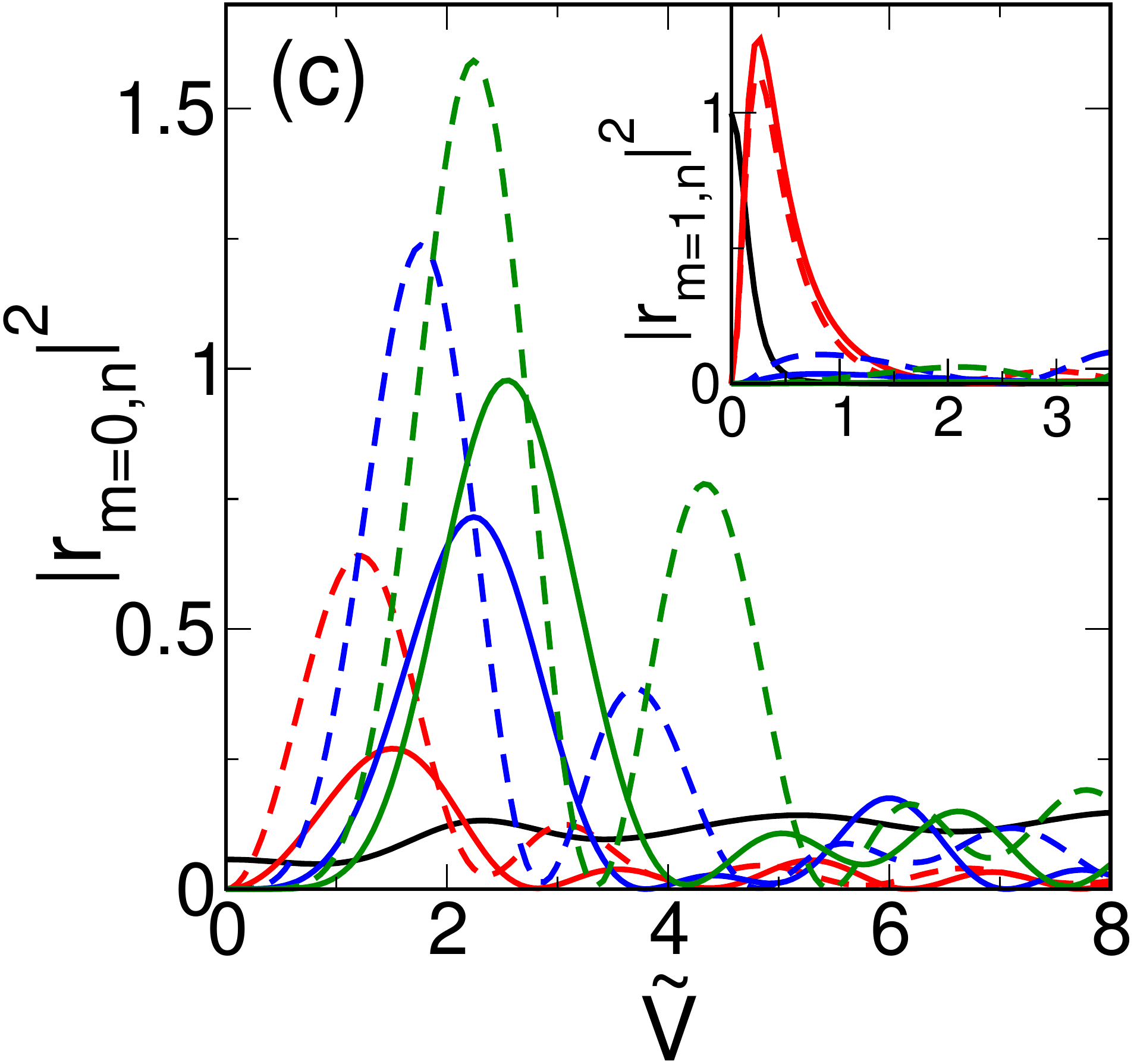}
\end{minipage}
\begin{minipage}{4.25cm}
\vspace{-0.05cm}
\includegraphics[width=4.1cm]{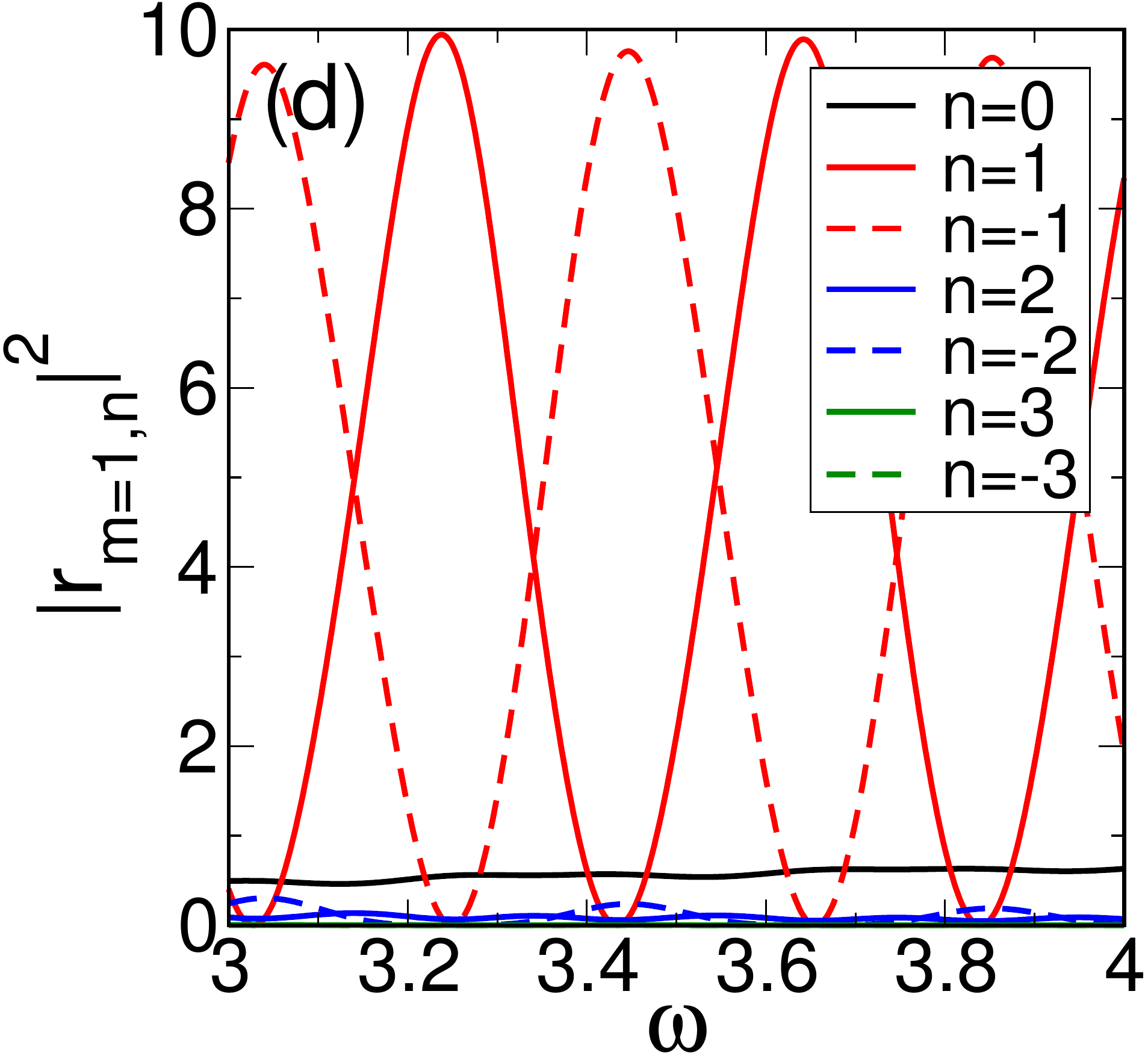}
\end{minipage}
\caption{
\label{pic3}(Color online) Intensity plot of the time-averaged far-field scattering efficiency $\bar{Q}$ of an  oscillating graphene quantum dot with finite bias ($V=1$), in dependence on $R$ and $\tilde{V}$ for $\omega=1$, $E=0.1$ [panel (a)], respectively, as a function $R$ and $\omega$ for  $\tilde{V}=0.75$ and $E=0.0629$ [panel (b)]. The lower figures give the corresponding (squared) amplitudes of the scattering coefficients $|r_{m,n}|^2$ as functions of $\tilde{V}$ [panel (c), where $E=0.1$, $\omega=1$, $R=4.5$ ] and $\omega$ [panel (d), where $E=0.0629$, $\tilde{V}=0.75$, $R=7.75$] of the reflected wave.}
\end{figure}

\subsection{Finite bias}
We finally consider an oscillating quantum dot with finite bias  ($V>0$), and allow for finite energies of the particle $E>0$. This case directly generalizes our previous work  for the static dot~\cite{HBF13a,SHF15a}.  Figure~\ref{pic3} gives the intensity of the time-averaged scattering efficiency in the far field. Keeping $\omega=1$ fixed, the scattering structures of the corresponding static  dot results as $\tilde{V}\to 0$, see panel (a). Here we observe relatively broad scattering signals   at $R=R_0\simeq 3$ and $R=R_0+k\pi$ ($k\in \mathbb{N}$) which can be attributed  to the $m$=0 mode, and further sharp resonances, belonging to modes with $m\geq 1$ 
[e.g., we find an $m$=1 resonance at $R\simeq 4.5$; cf. also the inset of panel (c)].  For larger values of $\tilde{V}$ resonances with higher energies ($n>0$) will be excited. Figure~\ref{pic3}~(c) indicates that now  a couple of superimposed $m$=0 states with different $n$ contribute. 
 As a result---compared to the dot with zero bias (see Fig.~\ref{pic1}~(a))---the scattering signals  are largely washed-out.  Panel~(b) shows the variation of the scattering efficiency with $\omega$. The strong signal observed at large $\omega$ around $R\simeq 7.75$ can be attributed to the resonance of a quasi-localized 
 $m$=1 mode. This mode appears also for the static dot with the same $V$, $R$, and $E$;  in the `anti-adiabatic' regime $\omega\gg 1$  the oscillation is so fast that the particle 
feels an averaged potential only.  Decreasing $\omega$ the intensity maximum oscillates with periodicity $\Delta \omega =\pi/R$ which is the difference between two subsequent resonances. The oscillation is  a consequence of the asymmetry of the dot potential ($V=1$), which shifts the locations of the resonances for the subsets of quantum dot energy levels with  $+|n|$ respectively $-|n|$. This shift is visualised in panel (d). We see that for the parameter set used partial waves 
 with $n=\pm 1$  give the dominant contributions to the scattering coefficients $r_{m=1,n}$. If the dot potential slowly oscillates  ($\omega< \tilde{V}$), 
 the situation changes dramatically. Then a large number of dot modes  are stimulated, with the result that the spectrum develops signatures composed of many partial waves with different $m,n$ [horizontal lines in the representation of panel (b)].

\section{Conclusions}
To summarize, we analyzed the scattering of a massless Dirac fermion, in the low-energy ($E$) sector, by a step-like cylindrical potential barrier ($U$), with and without bias ($V$), which harmonically oscillates in time with frequency ($\omega$) and  amplitude ($\tilde{V}$). In a sense, this setup models a gate-defined graphene quantum dot (but also charged impurities or short-ranged defects). Since the spatial extension of the gated region (radius $R$) is assumed to be on the scale of the wave length of the Dirac electron, quantum interference effects play an essential role. We note that single or double quantum dots of such size and arbitrary shape may be achieved in experiments by applying nanoscale topgates  on graphene nanoribbons or bilayer graphene~\cite{LHV10,AMY12,MKHWPS14}. To realize resonant scattering in small quantum dots with radii of, e.g., 100~nm,  we need---operating the dot at $E=3$~meV,  $V = 30$~meV, and $\tilde{V}  \sim$~100~meV---a relatively high potential modulation frequency of about 50 THz (which is the same order of magnitude as for an oscillating rectangular barrier~\cite{ZST08}).

Due to the chiral nature of the Dirac fermion Klein tunneling might be realized at normal and close to normal incidence, just as for the static dot.  Most notably, the oscillations of the quantum dot cause inelastic scattering, in contrast to what happens for the previously studied static circular potential barriers~\cite{CPP07,AU13,HBF13a,WF14}. In consequence, we observe dramatic changes in the scattering efficiency due to potential transitions into side-bands, having energies $E + n\hbar\omega$ ($n=0, \pm 1, \pm 2, \ldots$).  We showed that in the zero-bias case  ($E\simeq V \simeq 0$) forward scattering dominates and  observe a periodic emittance of Dirac electron waves by the quantum dot when resonance conditions are fulfilled. Thereby the particle is temporarily trapped in a vortex structure inside the gated dot region before it gets reemitted. Note that particle scattering is readily suppressed, however, by tuning, e.g., the bias, or the amplitude or frequency of the potential oscillation. In this way the gated quantum dot might act as a switch.  For a quantum dot with finite bias   ($E,\,V >0$), modes with finite angular orbital momentum appear and the energies belonging to side bands with positive and negative $n$ are shifted. As a result the time-averaged scattering efficiency  oscillates as the frequency is varied. The time-dependent phenomena detected in this work should be crucial for the analysis of field-driven transport through graphene  nanostructures, including the resonant transport through excited states~\cite{LHV10}, or for the design and control of  graphene-based quantum logic gates~\cite{ARP13}.
\appendix*
\section{Zero-frequency limit}
As an additional point, we demonstrate that the adiabatic limit (static quantum dot) is adequately reproduced within our approach.  In that, as $\omega \to 0$, the number of possibly excited energy levels more and more  increases,  the numerical treatment  becomes cumbersome. To proceed, we assume that the system realizes, for $\omega\to 0$,  at any infinitesimal point in time 
$\tau$, a static quantum dot with 
\begin{equation}\label{20}
Q^{\text{st}}[U]=\frac{4}{kR}\sum_{m}\left|r_{m}^{\text{st}}(U)\right|^{2}\,,
\end{equation}
where $U=V+\tilde{V}\sin \omega \tau $. Adding these contributions together over the period of oscillation $T$ gives the scattering efficiency in the adiabatic limit:
\begin{equation}\label{19}
\overline{Q}\left(\omega\rightarrow0\right)=\frac{1}{2\tilde{V}}\intop_{V-\tilde{V}}^{V+\tilde{V}}Q^{\text{st}}\left[U\right]\,\text{d}U\,.
\end{equation}

\begin{figure}[b]
\begin{minipage}{4.25cm}
\includegraphics[width=4.9cm,angle=-90]{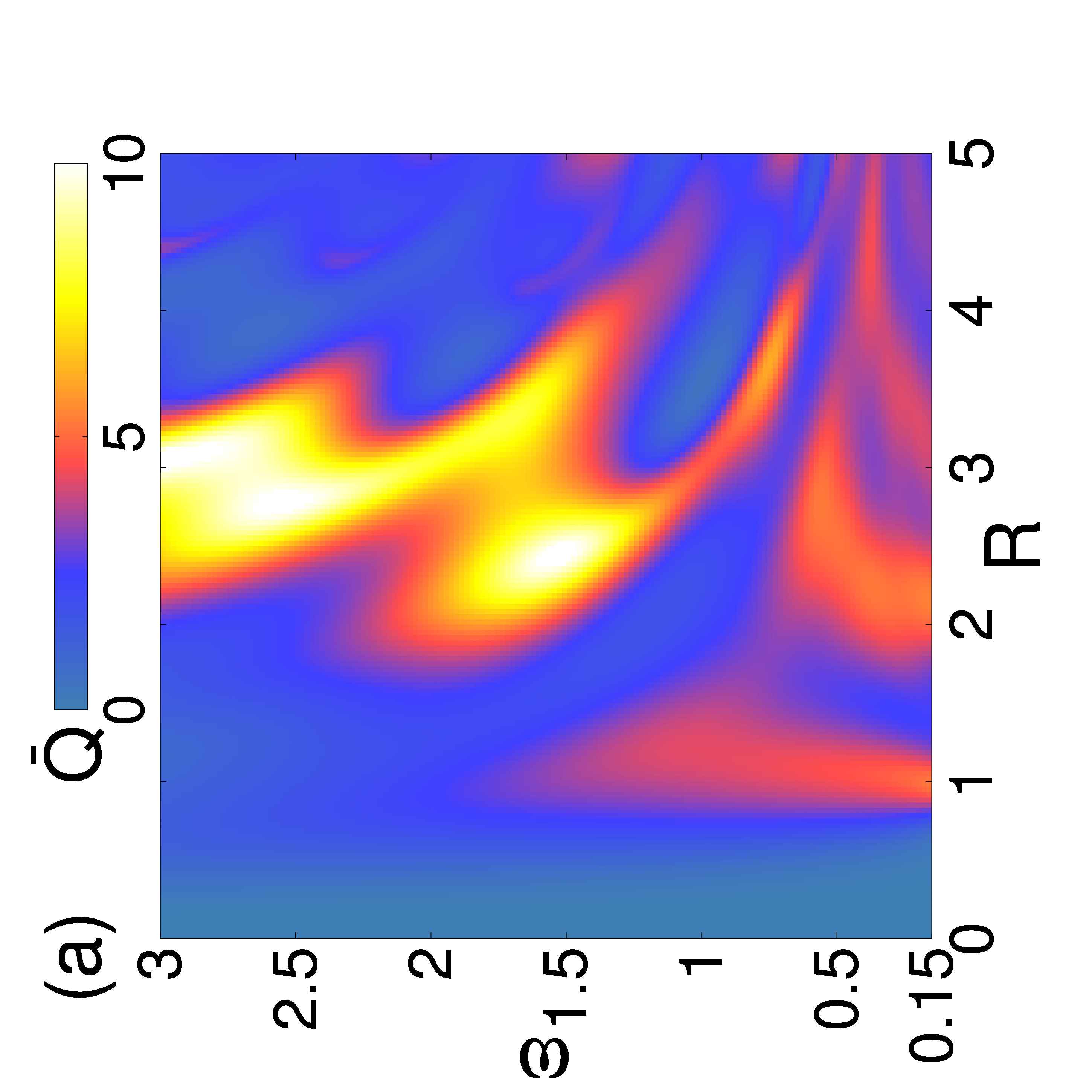}
\end{minipage}
\begin{minipage}{4.25cm}
\includegraphics[width=4.9cm,angle=-90]{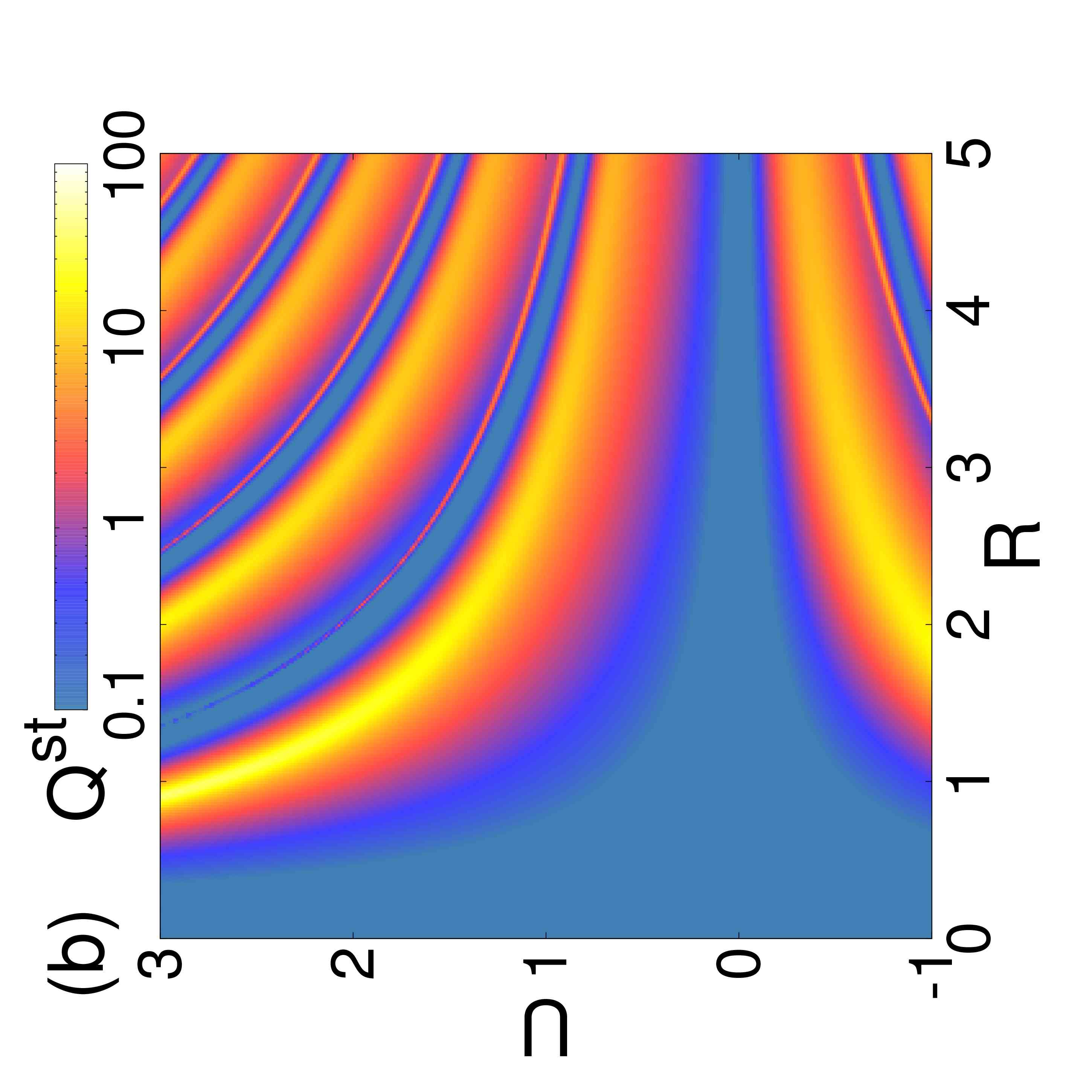}
\end{minipage}
\center
\vspace{-0.3cm}
\includegraphics[width=6.5cm]{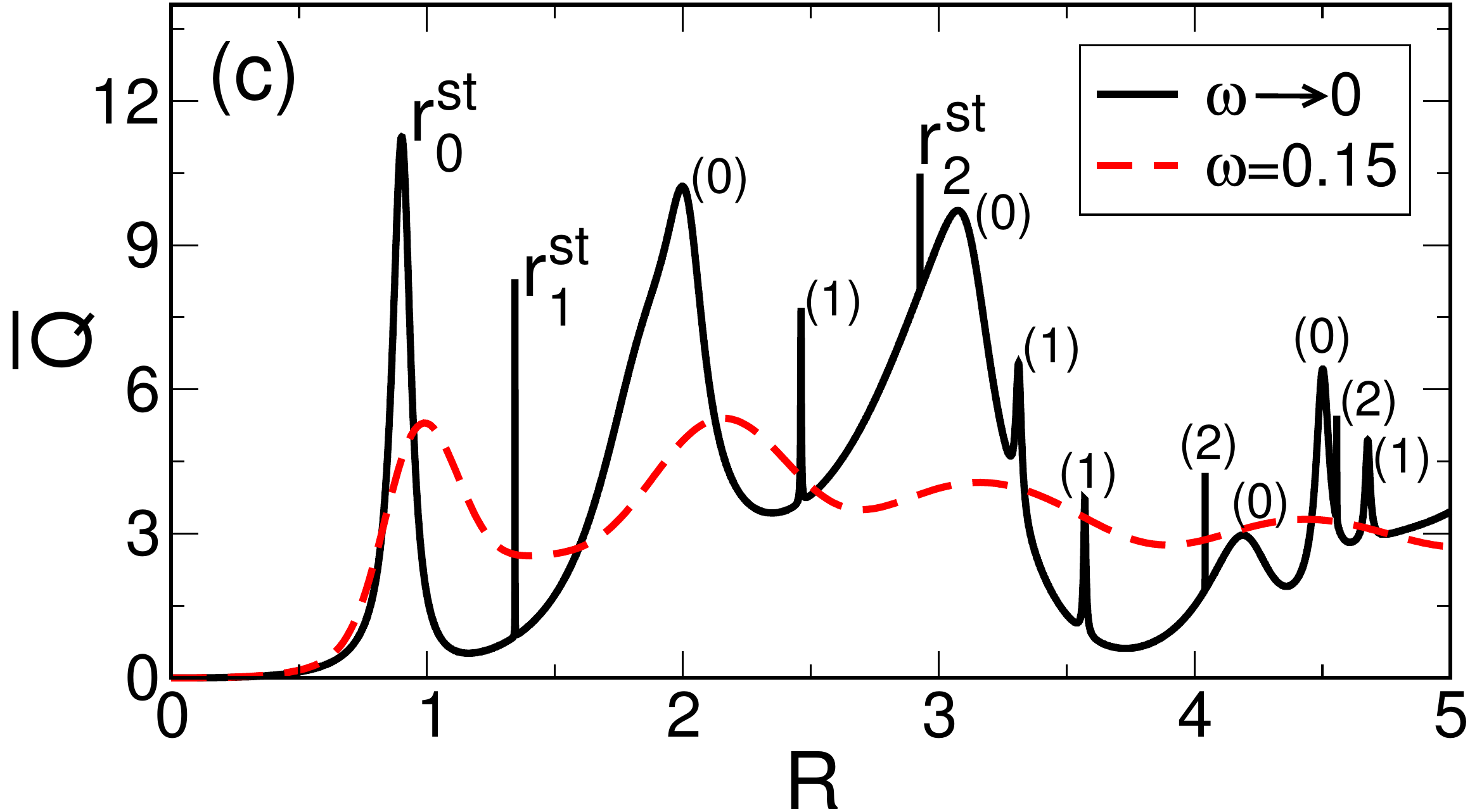}
\caption{
\label{pic4}(Color online) Scattering behavior approaching  the adiabatic limit. System parameters  are $E=0.1$, $V=1$, and $\tilde{V}=2$. Panel (a): Time-averaged scattering efficiency $\bar{Q}$ as  function of $R$ and $\omega$. Panel (b): static scattering efficiency $Q^{\text{st}}[U]$. Panel (c): time-averaged scattering efficiency  at fixed $\omega=0.15$ (red curve) compared to the result $\overline{Q}\left(\omega\rightarrow0\right)$ obtained in the adiabatic limit (black line). }
\end{figure}

Figure~\ref{pic4}~(a) illustrates the change in the intensity pattern of the  time-averaged scattering efficiency $\bar{Q}$ in the far-field as the oscillation frequency $\omega$ is reduced. For the parameter set used,  new resonances emerge below $\omega\lesssim 1$, which narrow and move to smaller $R$ values as $\omega\to 0$. These structures can be associated to the  static dot's $m$=0 mode and its overtones  [cf. panel~(c)]. In addition more localized   ($m=1, 2,\ldots$)  modes appear with less spectral weight, see the spikes in panel (c). Figure~\ref{pic4}~(b) gives the static scattering efficiency $Q^{\text{st}}[U]$, i.e., the integrand in Eq.~\eqref{19} in the range  $U=[V-\tilde{V}, V+\tilde{V}]$ at various $R$. The behavior of this quantity is mainly governed by the $|r_0^{\text{st}}|^2$ [broad (orange-to-yellow) signals] and $|r_0^{\text{st}}|^2$  [thin (red) curves] contributions.  

\acknowledgements{This work was supported by the Deutsche Forschungsgemeinschaft through Priority Program 1459 Graphene.}

\bibliography{ref}
\bibliographystyle{apsrev}

\end{document}